\documentclass{PoS}

\usepackage{amssymb}
\usepackage{amsmath}

\newcommand{\ba}[1]{\begin{array}{#1}}
\newcommand{\ea}{\end{array}}

\newcommand{\la}{\langle}
\newcommand{\ra}{\rangle}

\newcommand{\Dsymm}{{\ensuremath{\raisebox{0ex}[0ex][0ex]{\ensuremath{\overset{\leftrightarrow}{D}}}}}}
\newcommand{\mcA}{{\mathcal A}}
\newcommand{\mcO}{{\mathcal O}}
\newcommand{\MSbar}{{\overline{\text{MS}}}}

\title{Quark Contributions to Nucleon Momentum and Spin 
       from Domain Wall fermion calculations}

\ShortTitle{Quark momentum and spin from DW fermion calculations}

\author{\speaker{S.~N.~Syritsyn}\\
        Lawrence Berkeley National Laboratory,
        Berkeley, CA 94720, USA\\
        E-mail: \email{ssyritsyn@lbl.gov}}

\author{J.~R.~Green, J.~W.~Negele, A.~V.~Pochinsky\\
        Center\hspace{-0.0725pc} for\hspace{-0.0725pc} Theoretical\hspace{-0.0725pc}
Physics,\hspace{-0.0725pc} Massachusetts\hspace{-0.0725pc} Institute\hspace{-0.0725pc}
of\hspace{-0.0725pc} Technology,\hspace{-0.0725pc} 
        Cambridge,\hspace{-0.0725pc} MA\hspace{-0.0725pc} 02139,\hspace{-0.0725pc} USA\\
        E-mail: 
          \email{jrgreen@mit.edu},          
          \email{negele@mit.edu},
          \email{avp@mit.edu}}

\author{M.~Engelhardt\\
        Department of Physics, New Mexico State University, Las Cruces, NM 88003-8001, USA \\
        E-mail: \email{engel@physics.nmsu.edu}}

\author{Ph.~H\"agler\\
        Institut f\"ur Kernphysik, 
        Johannes Gutenberg-Universit\"at Mainz,\\
        Johann-Joachim-Becher-Weg 45,
        D-55128 Mainz, Germany \\
        E-mail:
          \email{haegler@kph.uni-mainz.de}}

\author{B.~Musch \\
        Thomas Jefferson National Accelerator Facility, Newport News, VA 23606, USA \\
        E-mail:
          \email{bmusch@jlab.org}}

\author{W.~Schroers\\
        NuAS, Stubenrauchstr. 3, 12357 Berlin\\
        E-mail: \email{Wolfram.Schroers@field-theory.org}}

\abstract{
  We report contributions to the nucleon spin and momentum from light quarks 
  calculated   using dynamical domain wall fermions with pion masses down to 
  $300\text{ MeV}$ and fine lattice spacing $a=0.084\text{ fm}$. 
  Albeit without disconnected diagrams, we observe that spin and orbital angular
  momenta of both $u$ and $d$ quarks are opposite, 
  almost canceling in the case of the $d$ quark, which agrees with previous 
  calculations using a mixed quark action.
  We also present the full momentum dependence of $n=2$ generalized form factors showing
  little variation with the pion mass.}

\FullConference{ The XXIX International Symposium on Lattice Field Theory - Lattice 2011\\
July 10-16, 2011\\
Squaw Valley, Lake Tahoe, California}

\begin{document}
\section{Introduction}
Nucleon structure calculations on a lattice are important for both testing QCD as the
fundamental theory of quarks and gluons and making predictions and complementing
experimental efforts that aim to measure the full three-dimensional picture of the proton and
nucleon, and to explore the origin of the nucleon spin. 
Lattice calculations of nucleon structure are pursued by many groups, 
see \cite{Pleiter:2011gw, Aoki:2010xg, Yamazaki:2009zq, Lenz:2009ar}
for recent advances.
In this work, we report results for quark momentum fraction, spin and orbital 
angular momentum from calculations with $N_f=2+1$ domain wall fermions.
We also report previous $N_f=2+1$ calculations using a mixed (domain wall valence 
and Asqtad sea) quark action~\cite{Hagler:2007xi} with substantially increased 
statistics and reduced errorbars~\cite{Bratt:2010jn}.

\begin{table}[ht!]
  \centering
  \caption{\label{tab:ensembles}Lattice ensembles used in this work. 
           Uncertainties in the pion masses are 
           dominated by the scale uncertainty.}
  \begin{tabular}{ccc|c|ccc|cc}
    \hline\hline
    $L_s^3\times L_t$ &
    $m_\pi\text{ [MeV]}$ &
    $m_\pi L_s$ &
      Sep. $T/a$ &
      $am_\text{sea}$ & 
      $am_\text{val}$ & 
      $am_\pi$ &
      \# confs &
      \# meas 
    \\ 
    \hline
    \multicolumn{9}{c}{Mixed action 
                       ($a=0.1240(25)\text{ fm}$, $am_\text{strange}=0.050$)}
    \\
    $20^3\times 64$ &  293(6) &  3.68 &  9 &
      0.007 & 0.0081 &  
      0.1842(7) &  
      463 & 3704
    \\
    $20^3\times 64$ &  356(7) &  4.48 &  9 &
      0.010 & 0.0138 &  
      0.2238(5) &  
      631 & 5048
    \\
    $28^3\times 64$ &  356(7) &  6.27 &  9 &
      0.010 & 0.0138 &  
      0.2238(5) & 
      274 & 2192
    \\
    $20^3\times 64$ &  495(10) &  6.23 &  9 &
      0.020 & 0.0313 &  
      0.3113(4) & 
      486 & 3888
    \\
    $20^3\times 64$ &  597(12) &  7.50 &  9 &
      0.030 & 0.0478 &  
      0.3752(5) &
      563 & 4504
    \\ 
    \hline
    \multicolumn{9}{c}{Unitary domain wall 
        ($a=0.0840(14)\text{ fm}$, $am_\text{strange}=0.030$)}
    \\
    $32^3\times64$ &  297(5) &  4.06 &  $12$ &
      0.004 & 0.004 & 
      0.1268(3) &
      616 & 4928 
    \\
    $32^3\times64$ &  355(6) &  4.86 &  $12$ &
      0.006 & 0.006 &  
      0.1519(3) &
      883 & 7064 
    \\
    $32^3\times64$ &  403(7) &  5.52 &  $12$ &
      0.008 & 0.008 & 
      0.1724(3) &
      528 & 4224
    \\
    \hline\hline
  \end{tabular}
\end{table}

We perform calculations that span different values of lattice spacing, volume and pion masses,
see Tab.~\ref{tab:ensembles}.
With the domain wall fermion action, we perform high-statistics calculations using
$300\text{ MeV}\lesssim m_\pi \lesssim 400\text{ MeV}$, and with the mixed action using
$300\text{ MeV}\lesssim m_\pi \lesssim 600\text{ MeV}$.

Quark momentum and angular momentum can be extracted from the matrix elements of the quark
energy-momentum tensor between nucleon states:
\begin{equation}
  \label{eqn:GFFn2}
  \begin{aligned}
  \la P^\prime |\bar{q}\big[\gamma^{\{\mu}\, i\Dsymm^{\nu\}} - \la\text{trace}\ra\big]q|P\ra 
    &
    = \bar{U}(P^\prime)\Big[A_{20}^q(Q^2)\gamma^{\{\mu} \bar{P}^{\nu\}} 
    + B_{20}^q(Q^2) \frac{i}{2M_N} \bar{P}^{\{\mu} \sigma^{\nu\}\alpha} q_\alpha 
    \\ &\quad\quad 
    + C_{20}^q(Q^2) \frac{1}{M_N} q^{\{\mu} q^{\nu\}} 
    - \la\text{trace}\ra\Big]U(P)
    ,
  \end{aligned} 
\end{equation}
where $q=P^\prime-P$, $\bar{P} = \frac12(P+P^\prime)$, $Q^2 = -q^2$ 
and $A_{20}$, $B_{20}$ and $C_2$ are momentum transfer-dependent 
\emph{$n=2$ generalised form factors} (GFF).
The forward value of $A_{20}(0)$ gives the quark momentum fraction inside the nucleon, 
while the combination of $A_{20}$ and $B_{20}$ gives the quark angular
momentum~\cite{Ji:1996ek}:
\begin{equation}
  \label{eqn:momfrac_angmom}
  \begin{aligned}
    \la x\ra &= A_{20}(0)\,, &&&
    \quad\quad\quad J_q &= \frac12\big[A_{20}(0) + B_{20}(0)\big].
  \end{aligned}
\end{equation}
Note that the $B_{20}$ GFF requires calculations with non-zero momentum transfer.
In this work we concentrate on $n=2$ GFFs, with $n=3$ GFFs reported
elsewhere~\cite{lhpc-gff2011-inprep}.

\section{Renormalization}

The energy-momentum operator in Eq.~(\ref{eqn:GFFn2}) is scale-dependent and its matrix elements
have to be renormalized for comparison with experiment. 
For the domain wall ensembles, we calculate the renormalization constants 
using the Rome-Southampton method~\cite{Martinelli:1994ty} 
following the RI${}^\prime$-MOM prescription.
In this scheme, the in- and out-quarks of the pertinent vertex function 
have the same off-shell 4-momentum $p=p^\prime$ 
which defines the scale of the renormalization point $\mu^2 = p^2$.
This scale must be in the window 
$\Lambda_\text{QCD}^2 \ll \mu^2 \ll \Lambda_\text{lat}^2 = a^{-2}$, 
i.e. in the perturbative region but small enough 
to avoid significant discretization errors.
There are indications that such a window does not exist for the hybrid action
ensembles~\cite{lhpc-gff2011-inprep}, and we use the renormalization constants 
from lattice perturbative calculations~\cite{Bojan-thesis}.
Since the gauge field in hybrid action calculations was HYP-smeared~\cite{Bratt:2010jn}, we
believe that the 1-loop perturbative calculations are a reasonably good approximation, 
and in Sec.~\ref{sec:results}, we will see close agreement between calculations using
perturbative and non-perturbative renormalization.
\begin{figure}[ht!]
  \centering
  \includegraphics[width=.6\textwidth]{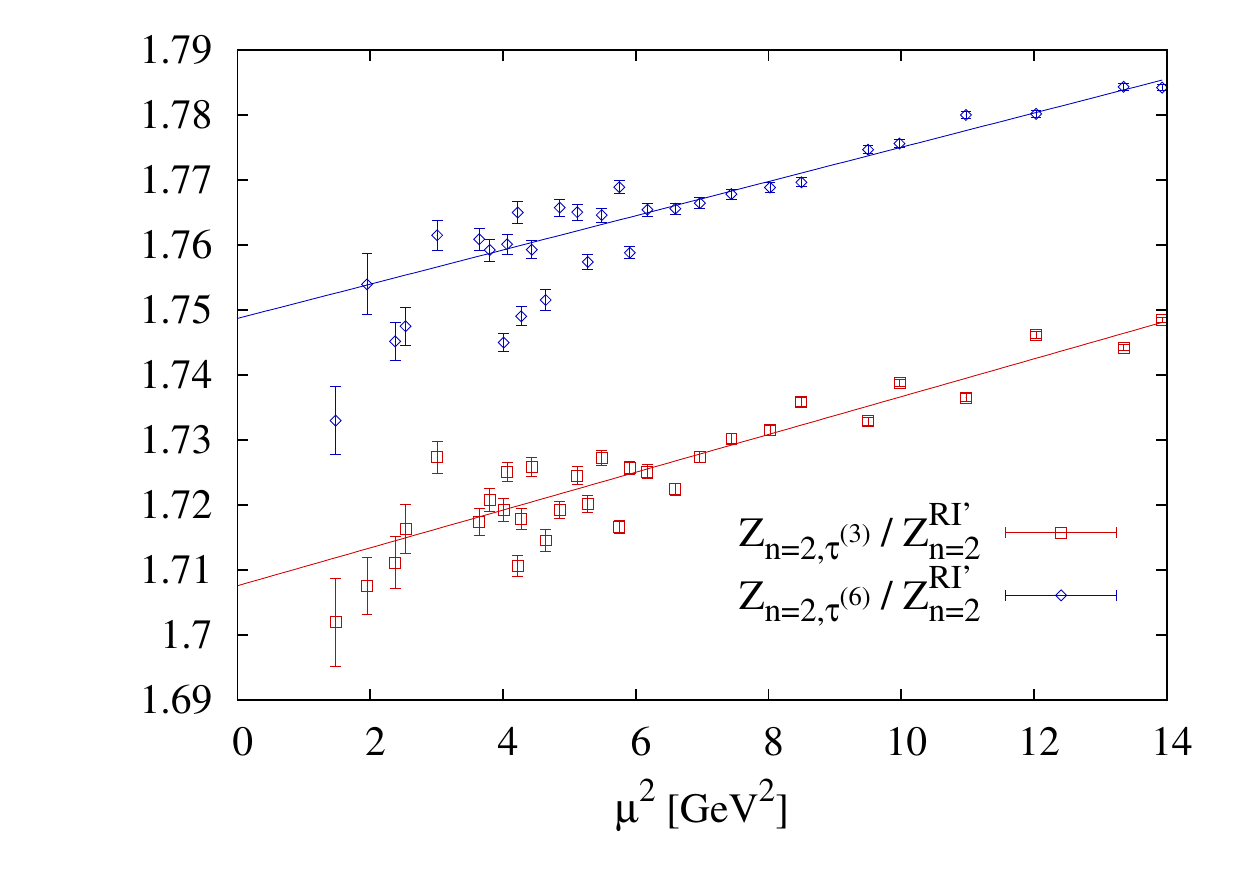}
  \caption{\label{fig:renorm-rank2}RI/MOM renormalization constant 
    divided by 3-loop perturbative running (\emph{scale-independent}).
    The extracted values must be multiplied with factors $Z_A/Z_\mcA$ and 
    $(Z^\MSbar_\mcO / Z^{\text{RI}^\prime}_\mcO)_{2\text{ GeV}}$ 
    to get the final renormalization constants.}
\end{figure}

We calculate vertex functions of the quark energy-momentum operator for two hypercubic
representations $\tau^{(3)}$ and $\tau^{(6)}$~\cite{Gockeler:1996mu} and the (local) 
axial-vector current operator $A_{x,\mu} = \bar{q}_x\gamma_\mu\gamma_5 q_x$.
We use the latter to exclude the quark field renormalization, employing the fact that the
\emph{conserved} axial-vector current $\mcA_{x,\mu}$ of the domain wall fermions, 
with good precision, should have trivial renormalization $Z_\mcA \approx1$, 
and calculating the ratio $Z_\mcA / Z_A$ from pion matrix elements separately.

In Fig.~\ref{fig:renorm-rank2} we show the ``scale-independent'' renormalization constants,
i.e., the lattice renormalization constants divided by 
the running constant $Z^{\text{RI}^\prime}(\mu)$ calculated perturbatively up to 3 loops 
in continuum theory~\cite{Gracey:2003yr}.
In the region $4 \lesssim \mu^2 \lesssim 14\text{ GeV}^2$, 
the ``scale-independent'' ratios vary within 
$\pm1\%$, demonstrating perfect matching between the lattice and the continuum 
renormalization factors. 
Assuming that the discretization errors contribute as $Z(a) = Z(0)+Z^\prime a^2\mu^2$, 
we extrapolate to $\mu=0$ and rescale our results to the final $\MSbar(2\text{ GeV})$ scheme. 
The final renormalization factors for the two representations are close to each other:
$Z^{\MSbar(2\text{ GeV})}_{\tau^{(3)}} = 1.708(4)$ and 
$Z^{\MSbar(2\text{ GeV})}_{\tau^{(6)}} = 1.749(2)$.

\section{\label{sec:results}Results}

To compute nucleon two- and three-point functions 
we use Gaussian-smeared sources tuned in Ref.~\cite{Syritsyn:2009mx}.
The source-sink separation $T$ for both calculations is roughly $1.0\ldots1.1\text{ fm}$ 
which we expect to be sufficient to suppress excited state effects at the precision level we
aim at.
We extract the matrix elements of the energy-momentum tensor using the standard 
ratio/plateau technique~\cite{Hagler:2003jd} and the overdetermined 
analysis~\cite{Syritsyn:2009mx,Hagler:2003jd}. 

\begin{figure}[ht!]
  \centering
  \begin{minipage}{.48\textwidth}
    \includegraphics[width=\textwidth]{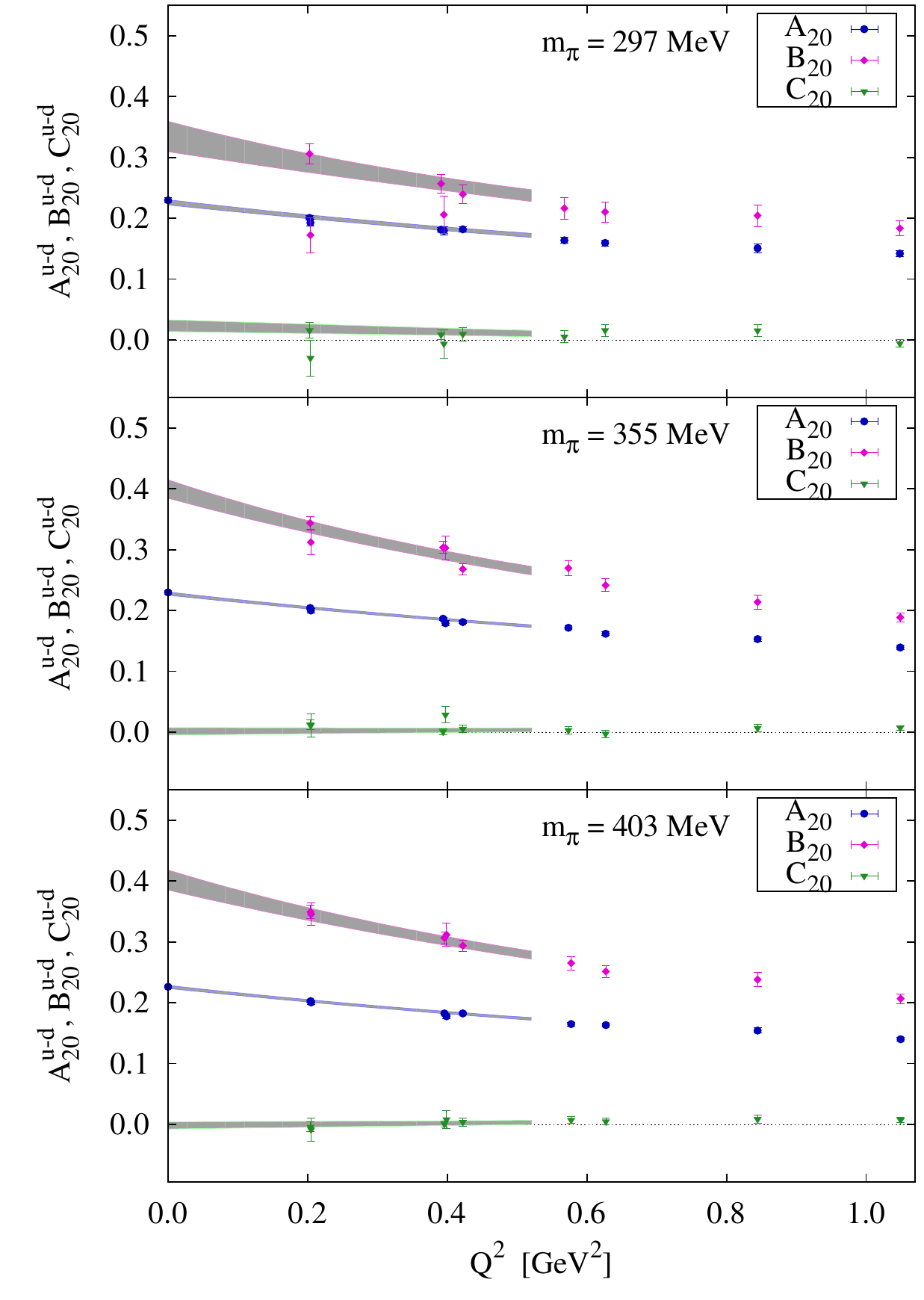}
  \end{minipage}~\hspace{.03\textwidth}~
  \begin{minipage}{.48\textwidth}
    \includegraphics[width=\textwidth]{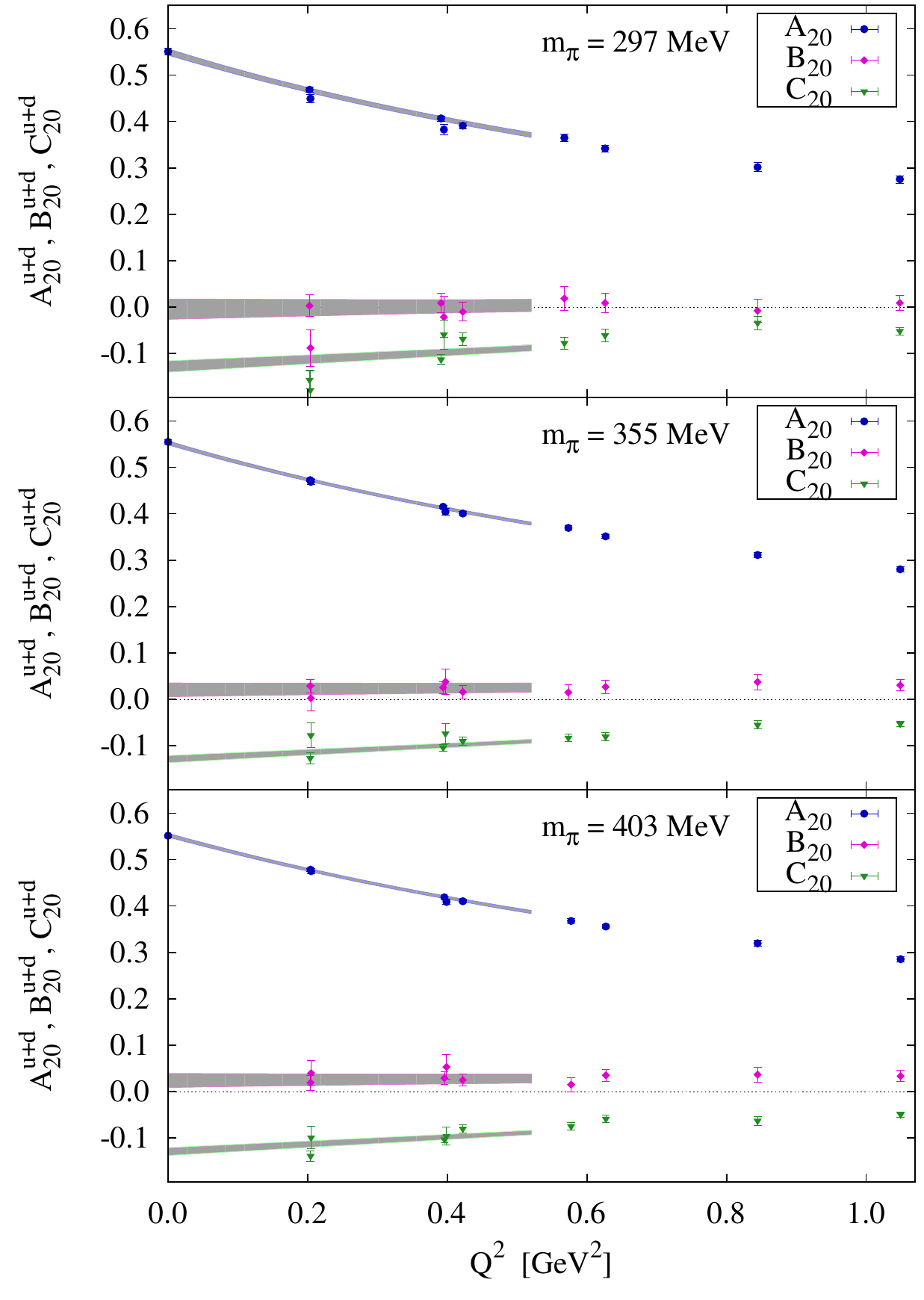}
  \end{minipage}
  \caption{\label{figs:ABC-vs-Q2}(Color online) 
    Isovector (left) and isoscalar (right) 
    $n=2$ Generalized form factors (GFF) of the proton 
    from domain wall calculations. }
\end{figure}

In Fig.~\ref{figs:ABC-vs-Q2} we show the dependence of GFFs on the momentum transfer.
The bands correspond to the dipole fits of each form factor and their statistical variation.
Using these fits, we extract the forward ($Q^2=0$) values of the isovector and isoscalar GFFs,
as well as their slopes.

It is remarkable that there are specific features of the GFFs independent of $m_\pi$, e.g. 
\begin{align}
|C_2^{u+d}| & \gg |C_{2}^{u-d}|  \approx 0,
&
|B_{20}^{u-d}| & \gg |B_{20}^{u+d}| \approx 0,
\end{align}
which are also consistent with previous calculations~\cite{Hagler:2007xi} and large-$N_c$
counting rules~\cite{Goeke:2001tz}.


\begin{figure}[ht!]
  \centering
  \begin{minipage}{.48\textwidth}
    \includegraphics[width=\textwidth]{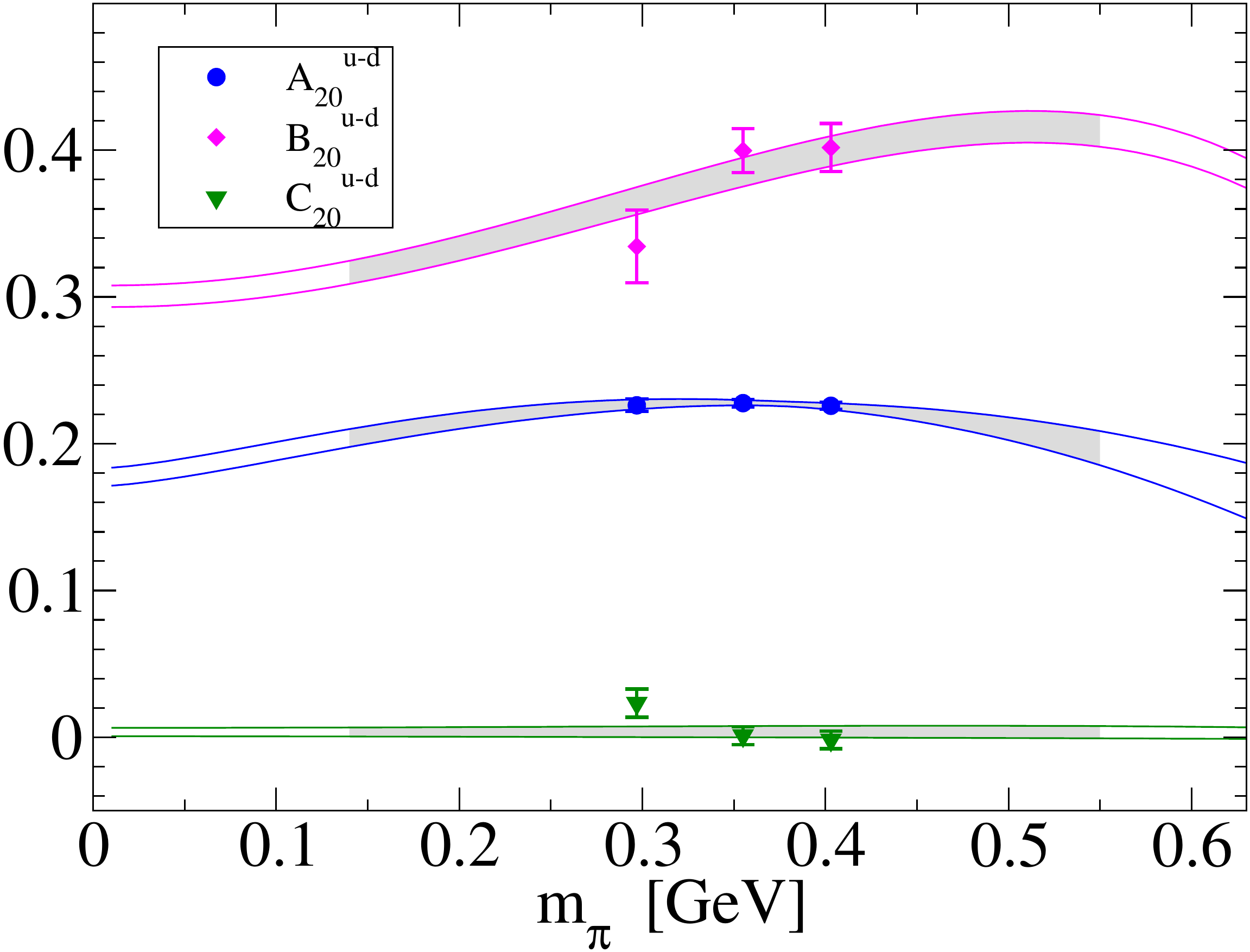}
  \end{minipage}~\hspace{.03\textwidth}~
  \begin{minipage}{.48\textwidth}
    \includegraphics[width=\textwidth]{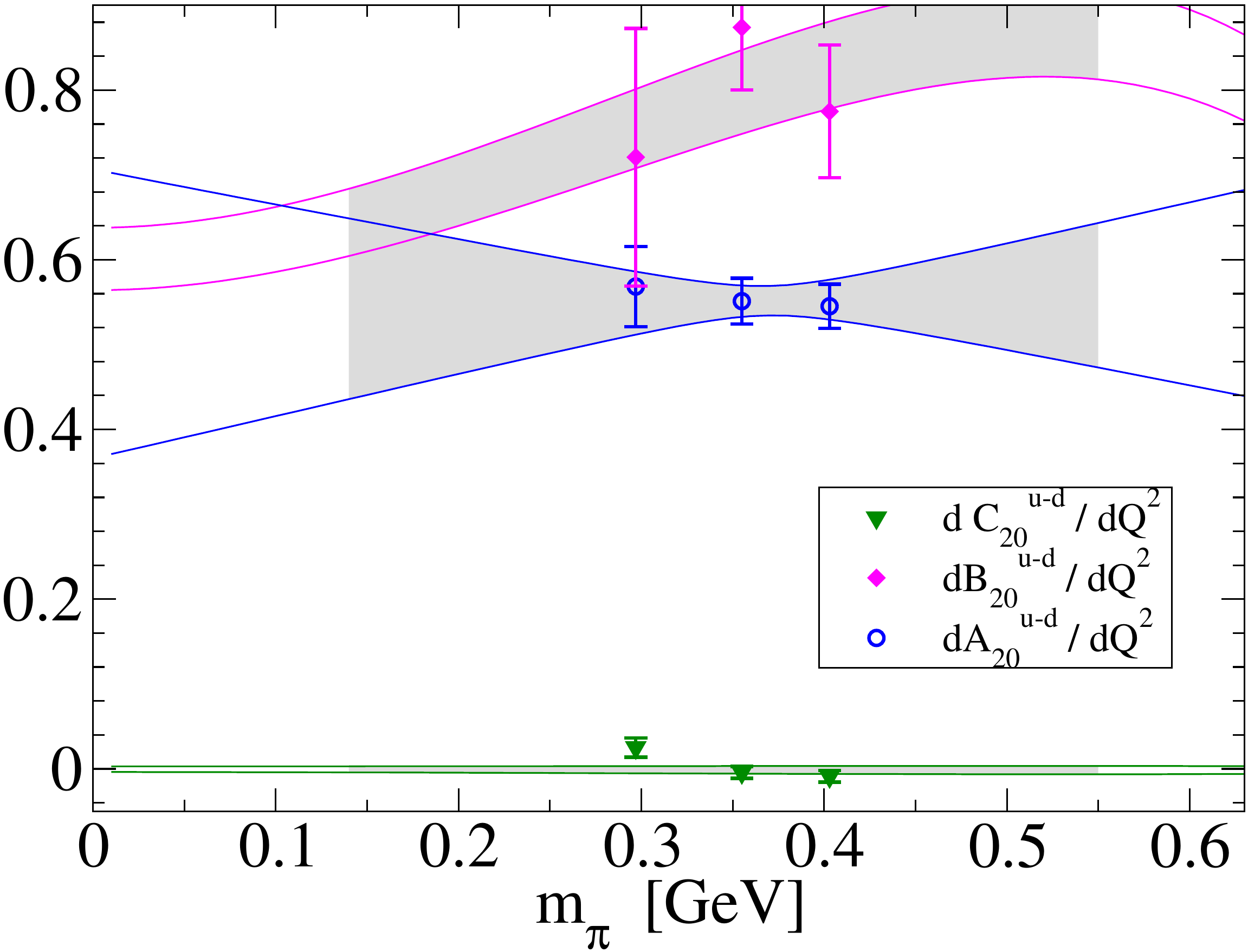}
  \end{minipage}
  \caption{\label{fig:chiral-extrap-isovec}(Color online)
    Chiral extrapolations of isovector GFFs 
    $A_{20}$, $B_{20}$, $C_{20}$ (left) and their derivatives $d/dQ^2$ (right) at $Q^2=0$.
    The ChPT predictions are taken from Ref.~\cite{Dorati:2007bk}.}
\end{figure}

We perform chiral extrapolation of the generalized form factors at $Q^2=0$ using 
baryon chiral perturbation theory~\cite{Dorati:2007bk}.
While the formulas in Ref.~\cite{Dorati:2007bk} predict both the $m_\pi$ and $Q^2$ dependence,
it is not clear if they are applicable for $Q^2\gtrsim0.2\text{ GeV}^2$.
We therefore restrict our analysis only to the forward values and their derivatives obtained
using the dipole extrapolation for $Q^2\to0$. 
In Fig.~\ref{fig:chiral-extrap-isovec} we show fitting and extrapolation results 
for the isovector GFFs.
The analysis of the isoscalar GFFs is similar.
The GFF derivatives help to constrain the fits considerably, 
although less than the GFFs themselves.


\begin{figure}[ht!]
  \centering
  \begin{minipage}{.48\textwidth}
    \includegraphics[width=\textwidth]{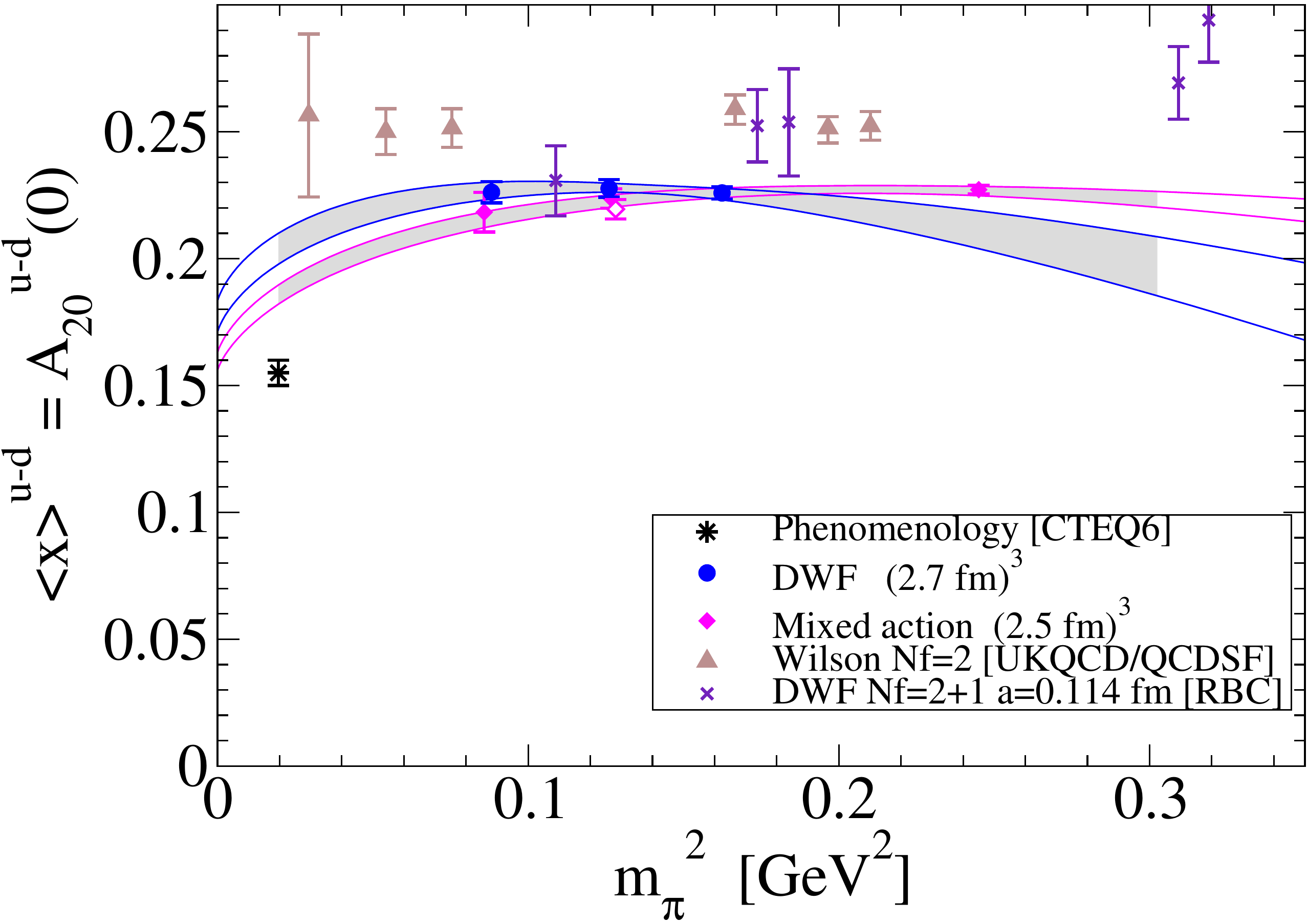}
  \end{minipage}~\hspace{.03\textwidth}~
  \begin{minipage}{.48\textwidth}
    \includegraphics[width=\textwidth]{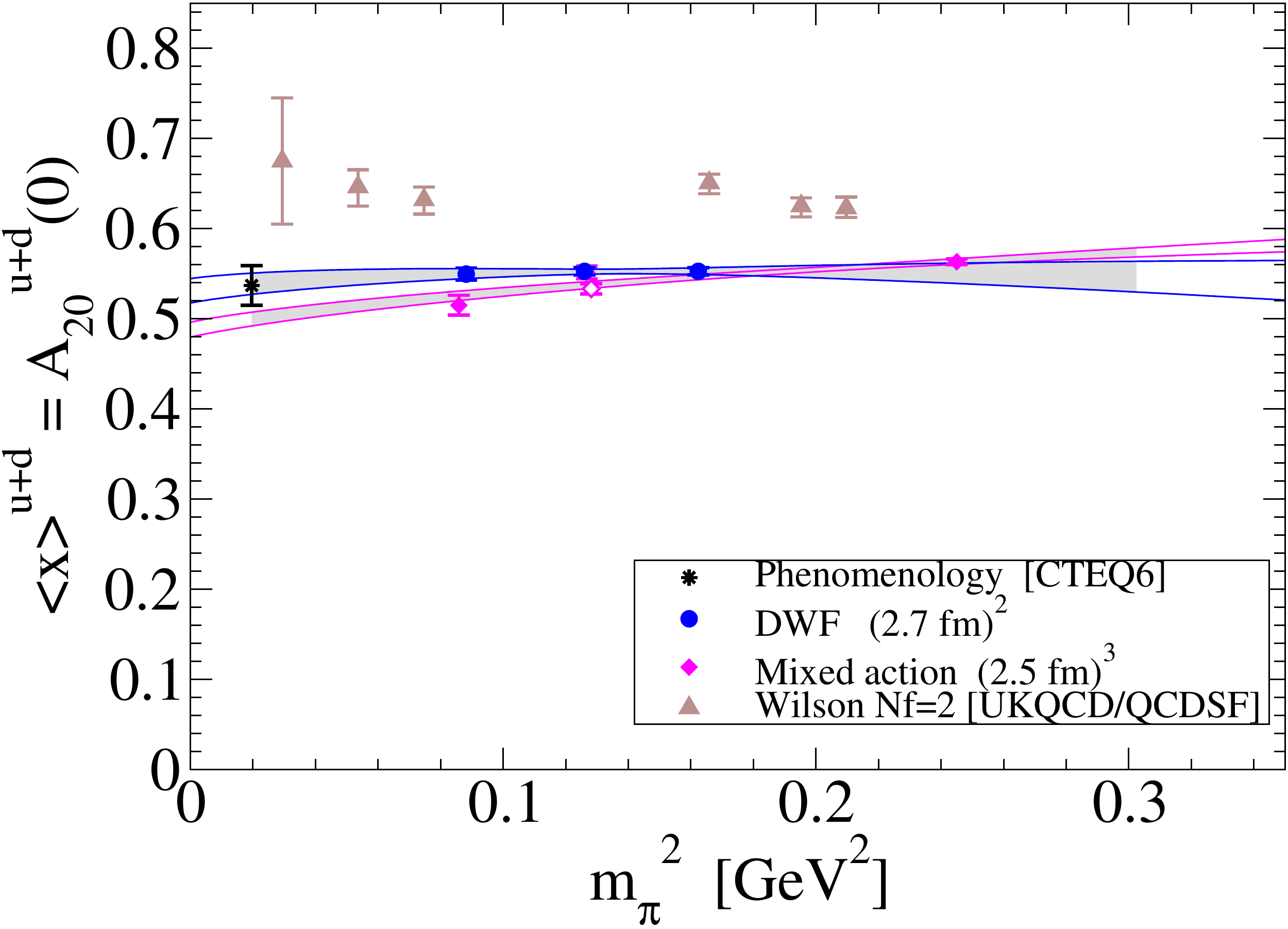}
  \end{minipage}
  \caption{\label{fig:momfrac}(Color online)
    Isovector (left) and isoscalar (right) quark momentum fraction in
    the proton compared to calculations in Ref.~\cite{Pleiter:2011gw, Aoki:2010xg}.}
\end{figure}

In Fig.~\ref{fig:momfrac} we show our results for the quark momentum fraction $\la x\ra$ for
both domain wall and hybrid action calculations.
It is remarkable that the results agree very well, although lattice spacings 
and renormalization procedures are different.
In addition we show the comparison of our results to those of Ref.~\cite{Pleiter:2011gw}
where $N_f=2$ Clover fermions were used and 
Ref.~\cite{Aoki:2010xg}, which used $N_f=2+1$ domain wall fermions with $a=0.114\text{ fm}$.
It is interesting that $N_f=2+1$ domain wall results tend to agree at smaller pion masses,
while $N_f=2$ Clover calculations disagree by $\approx15\%$.
Our chirally extrapolated value of $\la x\ra^{u-d}=0.204(6)$ significantly 
overshoots the phenomenological value $0.155(5)$. 
This disagreement can be the result of either unreliability of the chiral extrapolation or
presence of excited states in the matrix element calculation.
At the same time, the isoscalar momentum fraction $\la x\ra^{u+d}$ agrees well with the
phenomenology, albeit without disconnected contractions.
This fact may be an indication of the smallness of the disconnected contributions.

\begin{figure}[ht!]
  \centering
  \begin{minipage}{.50\textwidth}
    \includegraphics[width=\textwidth]{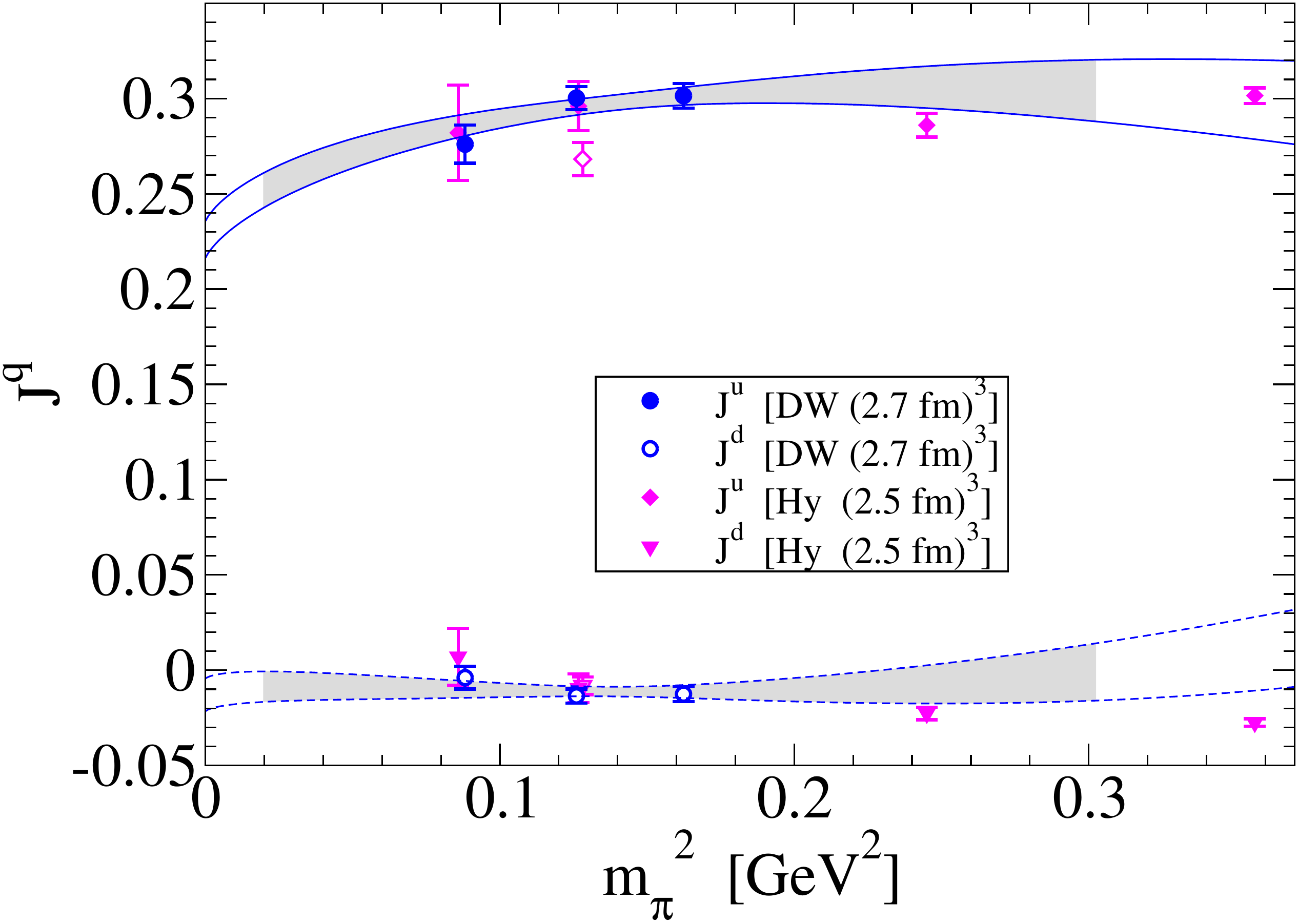}
  \end{minipage}~\hspace{.03\textwidth}~
  \begin{minipage}{.46\textwidth}
    \vspace{-.2cm}
    \includegraphics[width=\textwidth]{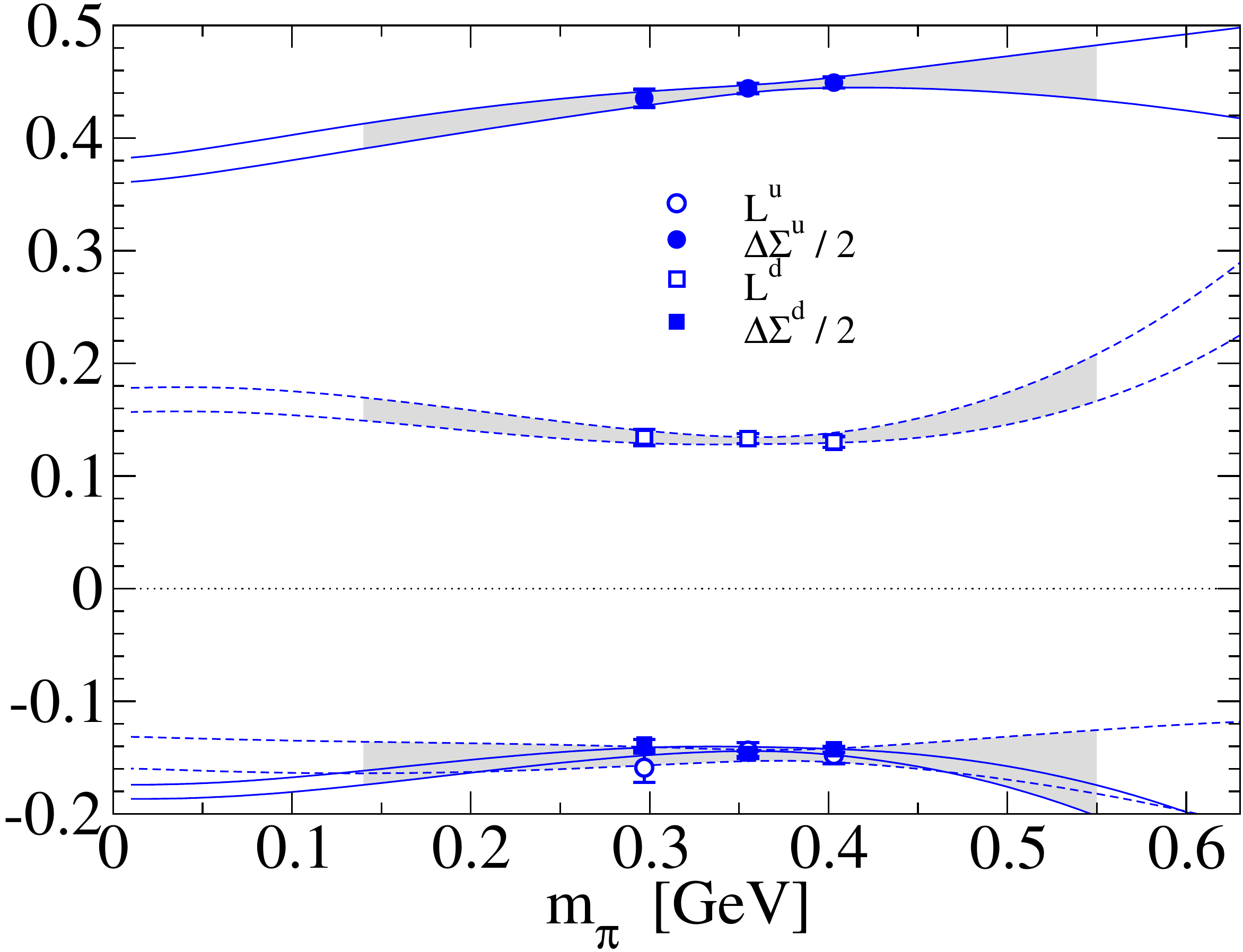}
  \end{minipage}
  \caption{\label{fig:quark-spin-oam}(Color online)
    Left: $u$ and $d$ quark contributions to the nucleon spin
    from the domain wall and hybrid action calculations. 
    Right: $u$ and $d$ quark spin and orbital
    momentum from the domain wall calculations.}
\end{figure}

In Fig.~\ref{fig:quark-spin-oam} we show quark angular momentum (disconnected contractions are
not included), as well as separate contributions of $u$ and $d$ quark spin and orbital angular
momentum (OAM) to the proton spin.
Although not shown, results from the hybrid action calculations~\cite{Bratt:2010jn} 
agree very well with the data in Fig.~\ref{fig:quark-spin-oam}(right).
The calculations with domain wall fermions confirm previous qualitative results\footnote{
  We do not include the disconnected contractions, which can potentially modify
  Eq.~\ref{eqn:UD_spin_oam_rel}.
}
that
\begin{align}
  \label{eqn:UD_spin_oam_rel}
  |J^d| &\ll |J^u|\,, & 
  |J^d| = |S^d + L^d| &\ll |S^d|,\,|L^d|\,, &
  |L^{u+d}| &\ll |L^u|,\,|L^d|\,.
\end{align}

\section{Discussion}
We have computed the generalized form factors of the nucleon in fully dynamical QCD
with $N_f=2+1$ flavors of chirally symmetric quarks. 
The pion masses $300-400\text{ MeV}$ are still too heavy to be realistic, 
however, they are arguably close enough for the applicability of chiral perturbation theory.
Nevertheless, we see disagreement between the extrapolated value of 
the isovector quark momentum fraction $\la x\ra^{u-d}$ and phenomenology.
The presented agreement with the previous calculations employing different quark action, 
lattice spacing and operator renormalization procedure indicates that the problem may reside 
in either insufficiency of the order of the chiral perturbation theory expansion
or presence of nucleon excited state contributions in the correlation functions. 
There are indications for excited state contributions in other calculations as
well~\cite{jrgreen-lat2011}.

The disagreement with the $N_f=2$ calculations may also indicate that the number of flavors
plays a crucial role for this quantity, leading to the conclusion that 
the low energy dynamics of QCD may affect $\la x\ra^{u-d}$ more than other quantities.

\section*{Acknowledgements}
This research was supported in part by funds provided by the U.S. Department of Energy (DOE)
under cooperative research agreement DE-FG02-94ER40818
and 
under Contract No. DE-AC02-05CH11231.
B.M. is supported by U.S. DOE contract No. DE-AC05-06OR23177.
M.E. is supported by U.S. DOE grant No. DE-FG02-96ER40965.
The U.S. Government retains a non-exclusive, paid-up, irrevocable,
world-wide license to publish or reproduce this manuscript for U.S. Government purposes.
Computer resources were provided by the DOE 
through the USQCD project at Jefferson Lab,
through its support of the MIT Blue Gene/L 
and through Argonne Leadership Computing Facility at ANL; 
and by the New Mexico Computing Applications Center (NMCAC) on Encanto.


\end{document}